\documentstyle[12pt,epsf]{article}

\def\Frac#1#2{\frac{\displaystyle{#1}}{\displaystyle{#2}}}
\def\lsim{\raise0.3ex\hbox{$\;<$\kern-0.75em\raise-1.1ex\hbox{$\sim\;$}}}
\def\gsim{\raise0.3ex\hbox{$\;>$\kern-0.75em\raise-1.1ex\hbox{$\sim\;$}}}

\newcommand{\beqn}{\begin{eqnarray}}
\newcommand{\beq}{\begin{equation}}
\newcommand{\eeqn}{\end{eqnarray}}
\newcommand{\eeq}{\end{equation}}

\begin{document}
\hyphenation{re-pha-sings bo-unds e-lec-tron u-ni-ver-sa-li-ty
sfer-mions in-de-pen-dent Yu-ka-wa se-arch char-gi-no phy-sics
spec-ci-fically su-per-sym-me-tric pha-ses model-in-de-pen-dent diago-na-lity 
                   an-ni-hi-la-tion pen-guin}
\begin{titlepage}
\begin{flushright}
FERMILAB-Pub-01/077-T\\
FTUV-01-05-29\\
IFIC-01-28
\end{flushright}
\vskip 1.5cm
\centerline {\Large{\bf Constraining models with vector-like fermions}} 
\vskip .5cm
\centerline {\Large{\bf from FCNC in $K$ and $B$ physics}}
 
\vskip 1cm
\centerline {G. Barenboim}
\vskip 0.2cm
\centerline {\it Theory Division, CERN, CH-1211 Geneva 23, Switzerland and}
\vskip 0.2cm
\centerline {\it FERMILAB Th. Group, MS 106 P.O.Box 500, 
Batavia IL 60510, USA}
\vskip 1cm
\centerline { F.J. Botella and O. Vives}
\vskip 0.2cm
\centerline {\it Departament de F\'{\i}sica Te\`orica and IFIC }
\vskip 0.2cm
\centerline {\it Universitat de Val\`encia-CSIC, E-46100, Burjassot, Spain}
\vskip 1cm 
\begin{abstract}
In this work, we update the constraints on tree level FCNC couplings
in the framework of a theory with $n$ isosinglet vector-like down
quarks. In this context, we emphasize the sensitivity of the $B \to
J/\psi K_S$ $CP$ asymmetry to the presence of new vector-like down
quarks.  This $CP$ asymmetry, together with the rare decays $B \to
X_{s,d} l \bar{l}$ and $K \to \pi \nu \bar{\nu}$ are the best options
to further constrain the FCNC tree level couplings or even to point
out, in the near future, the possible presence of vector-like quarks
in the low energy spectrum, as suggested by GUT theories or models of
large extra dimensions at the TeV scale.
\end{abstract}
\vskip 1cm

\vfill

\end{titlepage}

\newpage

\section{Introduction}
The study of flavor changing neutral currents (FCNC) in particle
physics phenomenology has played a key role in the advance of high
energy physics in the past decades. Already in the early seventies the
non-observation of FCNC was used to predict the existence of the charm
quark through the GIM mechanism \cite{GIM} well before its direct
experimental discovery. Subsequently, a precise prediction of its mass
was made from FCNC kaon processes \cite{gaillard}.  Later on, the
discovery of the bottom quark and the measure of the $B$--$\bar{B}$
mass difference indicated the presence of a heavy top quark.  Due to
this GIM mechanism, FCNC in the SM arise only at higher loop level and
are suppressed by powers of light quark masses and small mixing
angles.  This strong suppression make FCNC phenomena a privileged
ground to search for signs of new physics beyond the SM. In many
extensions of the SM there is no analog of the GIM mechanism to
protect FCNC processes and hence potentially large new physics
contributions can be expected. For instance, a minimal supersymmetric
extension of the SM with generic soft breaking terms gives rise to
dangerously large contributions to FCNC and $CP$ violating observables
through loop contributions with the supersymmetric particles running
in the loop \cite{susy}. Even more challenging (or perhaps more
dangerous) than these loop induced FCNCs is the inclusion of tree
level FCNC couplings.

Indeed, the presence of tree level FCNC is phenomenologically, a very
interesting possibility. A minimal extension of the SM with the only
addition of an extra isosinglet down quark in a vector like
representation of the SM gauge group induces FCNC couplings in the $Z$
and neutral Higgs boson couplings. Some, more realistic models,
include at least one vector-like quark per generation. These models
naturally arise, for instance, as the low-energy limit of an $E_{6}$
grand unified theory \cite{GUT}.  Moreover, vector-like quarks can appear in
models of large extra dimensions.  The existence of extra spatial
dimensions of the TeV size implies the existence of towers of Kaluza
Klein modes for the particles that propagate in the new (extra)
dimensions. Therefore, if some SM fermion propagates in the extra
dimensions, for each of these chiral quarks and leptons there is a
tower of vector-like fermions whose separation is of order $1/R$,
being $R$ the radius of the extra dimensions \cite{LED,FCLED}. From a
more phenomenological point of view, models with isosinglet quarks
provide the simplest self-consistent framework to study deviations of
$3\times 3$ unitarity of the CKM matrix as well as flavor changing
neutral currents at tree level. Of course, these FCNC couplings are
severely restricted by the low energy results on the different FCNC
processes.  Nevertheless, it is well known that even fulfilling these
strong constraints these couplings can have large effects on $B$
factory experiments on $CP$ violation \cite{vlq,BBBV,BBV}. In a recent
paper \cite{BBV}, we showed that the possible mismatch between the SM
expectations and the measured value for the $CP$ asymmetry in the $B
\to J/\psi K_s$ decay can be easily explained in a model with an
additional vector like down quark.  In this paper, we complete the
previous analysis and generalize it to the case where a tower of
vector-like down quarks (VLdQ) is present. In first place, we update
the strong low-energy constraints on the tree-level FCNC couplings and
then we concentrate on the $CP$ violation observables in the $B$
factories as the most efficient observables to find or at least
constrain these tree level FCNC couplings. Finally, we compare the
simplest case with a single VLdQ with models with several vector-like
down quarks.

\section{FCNC in the presence of isosinglet quarks}

As explained in the introduction, vector-like fermions appear in
different extensions of the SM, as for instance $E_6$ GUT models or,
remarkably enough, theories with large extra dimensions on the TeV
scale.  All these models have several new vector like fermions that
mix with the SM fermions and can have important effects on the low
energy FCNC phenomenology. In this paper, we will study FCNC processes
with external down quarks, and so we will be mainly interested in
models with isosinglet vector-like down quarks (VLdQ). Note that the
presence of extra vector-like down (up) quarks generates FCNC in the
down (up) couplings to the $Z$ and $H$. In view of these new
possibilities of having VLdQs in the low energy spectrum, and
especially, to reach a rough understanding of FCNC effects in models
with large extra dimensions \cite{LED,FCLED}, we present our bounds in
a general framework with $n$ additional VLdQs. A general analysis with
both up and down isosinglets and isodoublets will be presented
elsewhere \cite{WIP}.

From the low energy point of view, the model we analyze here has been
thoroughly described in Ref.~\cite{vlq,LuisJoao}. Its main feature is
that the additional VLdQs mix with the three ordinary quarks and
consequently the mass matrix in the down sector is now $(3+n)\times
(3+n)$.  This matrix is diagonalized by two $(3+n)\times(3+n)$ mixing
matrices and only the left-handed rotation is observable giving rise
to the CKM matrix. As we have already announced in the introduction,
the fingerprint of this model is that it allows tree level FCNC in the
$Z$ and $H$ vertices.  To see how these FCNC couplings appear, we can
work in the basis where the up quark mass matrix is diagonal. Here the
down quark mass matrix is diagonalized by a $(3+n)\times (3+n)$
unitary matrix, $V_{\alpha \beta}$.  In this model charged current
couplings are unchanged except that $V_{CKM}$ is now the $3 \times
(3+n)$ upper submatrix of $V$, and at low energies a non-unitary
$3\times 3$ mixing matrix appears at tree level.  However, the mixing
of doublet and singlet weak eigenstates modifies the neutral current
couplings that in terms of flavor eigenstates will be
$V^\dagger \cdot \mbox{Diag.}(1,1,1,0) \cdot V \neq
\mbox{I}_{4\times4}$. So, neutral current interactions are given by,
\begin{eqnarray}
&{\cal L}_Z = \Frac{g}{2 \cos \theta_W} \left[\bar{u}_{L i}\ \gamma^\mu\ 
u_{L i}\ -\  \bar{d}_{L \alpha}\  U_{\alpha \beta}\ \gamma^\mu\ d_{L \beta}\ - 
\nonumber \right.\\
& \left.\ 2\ \sin^2 \theta_W\ J^\mu_{em} \right]\ Z_{\mu}, \nonumber \\
&{\cal L}_H = \Frac{g}{2 M_W} \left[\bar{u}_{L i}\ m_i^u u_{L i}\ -\  
\bar{d}_{L \alpha}\  U_{\alpha \beta}\ m_\beta^d\ d_{L \beta}\right] \nonumber
\\
&U_{\alpha \beta }=\sum_{i=1}^3V_{i \alpha}^{*}V_{i\beta},
\label{ZFCNC}
\end{eqnarray}
where $U_{ds}$, $U_{bs}$ or $U_{bd} \neq 0$ would
signal new physics and the presence of FCNC at tree level.

In the following, we analyze the effects of these new couplings in
FCNC processes and we update the phenomenological bounds on them. More
specifically, in processes changing flavor in one unit, $\Delta F=1$,
that in the SM go through electroweak penguin diagrams of order $G_F\,
\alpha\, V_{t i}^* V_{t j}$, we simply consider the dominant tree
level FCNC contributions, order $G_F U_{i j}$.  Similarly in $\Delta
F=2$ processes mediated by boxes in the SM of order $G_F\, \alpha
\,(V_{t i}^* V_{t j})^2$ we include two different additional
contributions, a pure tree level diagram with two FCNC vertices, order
$G_F U_{i j}^2$ and a mixed SM vertex contribution with a tree level
vertex, roughly order $G_F\, \alpha\, V_{t i}^* V_{t j}\, U_{i j}$
\cite{BB}.  Other new physics contributions in this framework will
always be suppressed by additional loop factors or higher powers of
the FCNC couplings with respect to the contributions considered
here\cite{BotellaRoldan}.  Therefore the effective low energy theory
we consider is then identical to the SM with a non-unitary CKM matrix
and the presence of tree level FCNC as shown in Eq.~(\ref{ZFCNC}).

\section{Experimental constraints on the extended mixing matrix}

As we have seen in the previous section, all the new physics effects in 
our model are encoded in the extended mixing matrix that gives 
rise to tree level FCNC in the $Z$ and $H$ vertices.
The minimal extension from the SM would consist in the addition of a single
vector-like down quark. In this situation the mixing matrix can be 
parametrized as \cite{quico},
\begin{equation}
V=R_{34}\left( \theta _{34},0\right) R_{24}\left( \theta _{24},\phi
_{3}\right) R_{14}\left( \theta _{14},\phi _{2}\right) V_{CKM}^{SM}
\left( \theta _{12},\theta _{13},\theta _{23},\phi_{1}\right),
\label{matrixV}
\end{equation}
where $V_{CKM}^{SM}\left( \theta _{12},\theta _{13},\theta _{23},\phi
_{1}\right) $ is $4\times 4$ block diagonal matrix composed of the standard
CKM \cite{ChauKeung,PDG} and a $1 \times 1$ identity in the $(4,4)$
element, and $R_{ij}\left( \theta _{ij},\phi _{k}\right) $ is a
complex rotation between the $i$ and $j$ ``families''. Note
that, in the limit of small new angles, we follow the usual phase 
conventions. In fact, this $4 \times 4$ mixing matrix can represent a good
approximation to more complete models with several vector-like generations
if the mixings are hierarchical, similarly to the standard CKM matrix. 

On the other hand, given that the deviations from unitarity of the 
CKM mixing matrix are experimentally known to be small, it is possible to
use an approximate parametrization of the mixing matrix valid for
an arbitrary number of vector-like quarks \cite{LuisJoao}. This is an 
extension of the usual Wolfenstein parametrization of the CKM 
\cite{wolfenstein} in terms of, 
\begin{eqnarray}
\label{param}
V_{us}= \lambda\ \ \ \ \ &U_{d d}=\sum_{i=u,c,t}|V_{i d}|^2=1-D^2_d\
\ &
U_{s d}=\sum_{i=u,c,t}V_{i s}^{*}V_{i d}\ \
\nonumber \\
V_{cb}= A \lambda^2\ \ \ \ &
U_{s s}=\sum_{i=u,c,t} |V_{i s}|^2=1-D^2_s \ \ &
U_{b s}=\sum_{i=u,c,t}V_{i b}^{*}V_{i s}\ \
\nonumber \\
V_{ub}= A \mu\lambda^3 e^{i\phi}\ \ &
U_{b b}=\sum_{i=u,c,t}|V_{i b}|^2=1-D^2_b \ \ &
U_{b d}=\sum_{i=u,c,t}V_{i b}^{*}V_{i d}\ \ 
\end{eqnarray}
with $(U_{sd},U_{bd},U_{bs})$ general complex numbers and 
$D^2_j=\sum_{i=4}^{n+4} |V_{i j}|^2$ positive real numbers.
It is possible to obtain the remaining elements of the 
$3\times3$ submatrix corresponding to the SM mixing matrix as a function of
($\lambda$, $A$, $\mu$, $\phi$, $D^2_d$, $D^2_s$, $D^2_b$, $U_{sd}$, 
$U_{bd}$, $U_{bs}$). In fact, taking 
into account that ($D^2_d$, $D^2_s$, $D^2_b$) can be at most of order 
$\lambda^3$ and ($U_{sd}$, $U_{bd}$, $U_{bs}$) are experimentally constrained 
to be $\leq {\cal O}(\lambda^4)$, we can keep quadratic or linear terms in 
($D^2_d$, $D^2_s$, $D^2_b$) and ($U_{sd}$, $U_{bd}$, $U_{bs}$) respectively 
and we obtain to  ${\cal O}(\lambda^6)$,
\begin{eqnarray}
\label{wolf4}
V_{ud}&=& 1 - \Frac{\lambda^2}{2} - \Frac{\lambda^4}{8}  - 
\Frac{1 + 8 A^2 \mu^2}{16} \lambda^6 - \Frac{D^2_d}{2} + 
\Frac{D^2_d - 2 D^2_s}{4} \lambda^2 - \nonumber \\&&
 \Frac{D_d^4}{8} + \lambda \mbox{Re}\{U_{sd}\} + {\cal O}(\lambda^7) 
\nonumber \\
V_{cs}&=& 1 - \Frac{\lambda^2}{2} - 
\Frac{1+4 A^2}{8} \lambda^4 + \Frac{1 - 4 A^2 +16 A^2 \mu \cos \phi}{16} 
\lambda^6 \nonumber -\\&& \Frac{D^2_s}{2}- \Frac{D^2_s}{4} 
\lambda^2 +A \lambda^2 \mbox{Re}\{U_{bs}\} -\Frac{D_s^4}{8}  + 
{\cal O}(\lambda^7) \nonumber \\
V_{tb}&=& 1 -  \Frac{A^2}{2} \lambda^4  -\Frac{A^2 \mu^2}{2}
\lambda^6 - \Frac{D^2_b}{2} -\Frac{D_s^4}{8}+ 
{\cal O}(\lambda^7) \nonumber \\
V_{cd}&=& -\lambda + \left(\Frac{A^2}{2}-A^2\mu e^{-i \phi}\right) \lambda^5 
+ \Frac{D^2_d-D^2_s}{2} \lambda +U_{sd}+
\nonumber \\
&& \left(U_{bd} A - \Frac{U_{sd}^*}{2}\right) \lambda^2  + 
{\cal O}(\lambda^7) \nonumber \\
V_{td}&=& A(1-\mu e^{-i \phi})\lambda^3  + 
\Frac{A}{2}\mu e^{-i \phi} \lambda^5 +\Frac{A}{2}D^2_s\lambda^3 + U_{bd} +\nonumber \\&& 
\Frac{A}{2}\left(D^2_b-D^2_d\right)\left(1-\mu e^{-i \phi}\right) \lambda^3 - A \lambda^2 U_{sd}
+ {\cal O}(\lambda^7) \nonumber \\
V_{ts}&=& -A \lambda^2 + \left(\Frac{A}{2} -A\mu e^{-i \phi}\right)
 \lambda^4+ U_{bs}+ \Frac{A}{2}\left(D^2_s-D^2_b\right)\lambda^2
+ \nonumber \\&&
\Frac{A}{8} \lambda^6 + {\cal O}(\lambda^7)
\end{eqnarray}
This is a completely general parametrization of the $3\times3$ submatrix of
a complete $(3+n)\times(3+n)$ unitary matrix\footnote{Notice that this 
parametrization is approximate to ${\cal O}(\lambda^6)$ but an exact 
solution can be obtained numerically \cite{LuisJoao}}.
Hence, it includes also the simplest case of a single vector-like quark.
In this last case, there are several relations among the 13 parameters of this 
general matrix and only the nine independent parameters of a unitary
$4\times4$ mixing matrix remain. In fact we have, 
\begin{eqnarray}
\label{min4}
D_d^2=|V_{4 d}|^2,\ \ \ \ &  D_s^2=|V_{4 s}|^2 , \ \ \ \ & D_b^2=|V_{4 b}|^2 , 
\nonumber \\
U_{sd}=-V_{4 s}^* V_{4 d} , \ \ \ \ & U_{bs}=-V_{4 b}^* V_{4 s} ,\ \ \ \ &
 U_{bd}=-V_{4 b}^* V_{4 d}. 
\end{eqnarray}
Hence, we have 4 relations among the parameters in 
Eq.~(\ref{param}),
\begin{eqnarray}
\label{relation}
|U_{sd}|^2=D_s^2 D_d^2\ \ \ \ &|U_{bs}|^2=D_b^2 D_s^2 \nonumber \\ 
|U_{bd}|^2=D_b^2 D_d^2\ \ \ \  & \ \ \ \ \ \ \ \ \ \  U_{sd} U_{bs} U_{bd}^* = D_b^2 D_s^2 D_d^2 .
\end{eqnarray}
In the general case these equalities are replaced by inequalities,
\begin{eqnarray}
\label{ineq}
|U_{sd}|^2\leq D_s^2 D_d^2\ \ \ \ &|U_{bs}|^2\leq D_b^2 D_s^2 \nonumber \\ 
|U_{bd}|^2\leq D_b^2 D_d^2\ \ \ \ &
\end{eqnarray} 
 
Naturally, the elements of the extended mixing matrix corresponding to
CKM elements are experimentally measured in low energy
experiments. Some of these measurements are obtained from tree-level
processes, therefore not affected by new physics to a high degree of
approximation. Hence they constrain directly the corresponding
elements of our extended mixing matrix. Specifically, we have
\cite{PDG,vcs}, at $95\%$ C.L.,
\begin{itemize} 
\item $0.2150 < |V_{us}| = \lambda < 0.2242$,
\item $0.0364 < |V_{cb}| = A \lambda^2 < 0.0440$,
\item $0.074 < |V_{ub}/V_{cb}|=\mu \lambda < 0.106$,
\item $0.9719 < |V_{ud}| < 0.9751$,
\item $0.192 < |V_{cd}| < 0.256$,
\item $0.948 < |V_{cs}| < 1.0$.
\end{itemize}
These constraints restrict the different parameters in Eq.~(\ref{wolf4}), and
so from the bounds on $|V_{ud}|$ and $|V_{cs}|$, we obtain 
$D_d^2 \leq 5.5 \times 10^{-3} $ and  $D_s^2 \leq 5.6 \times 10^{-2}$ 
respectively. 
Another constraint \cite{vlq,LuisJoao,Paco} comes from the $SU(2)_L$ coupling
of the $Z^0$ to $b\overline{b}$. In the SM, this coupling is 
$(V_{CKM}^\dagger \cdot V_{CKM})_{bb}= 1$, but in this model it is modified to 
$U_{bb}$; hence, we have \cite{Paco} $D_b^2 \leq 9 \times 10^{-3}$. 

Notice that, in the general case, the $D_i^2$ parameters are
completely independent from the FCNC couplings $U_{\alpha
\beta}$. However, in a more definite model, as for instance the single
vector-like quark model, due to the unitarity of the $4 \times 4$
matrix, these constraints have a strong impact on all other elements
of the extended mixing matrix and consequently also on the tree level
FCNC couplings, as shown in Eq.~(\ref{relation}). In any case, even
with these restrictions, the FCNC couplings can have large effects on
rare processes, where the SM contributes at the 1 loop level and new
physics is allowed to compete on equal footing.  In fact, most of the
experimental results are well accommodated within a pure 3 generations
SM and hence these measurements provide additional constraints on the
FCNC couplings.  In other observables, like $B^0$ $CP$ asymmetries or
$B$ rare decays, the experimental results may still differ from the SM
prediction once the experimental accuracy is increased in the near
future. Therefore, we analyze these observables from a slightly
different point of view, and in the last section we show what are the
possible deviations from the SM consistent with the constraints
discussed here.

In the first place we analyze the constraints associated with kaon 
physics and then the constraints we get from the B system. 

\subsection{Kaon physics constraints}

Rare kaon decays and $CP$ violating observables in the kaon sector can receive 
important contributions from the FCNC coupling $U_{sd}$. In fact, this new
coupling is constrained mainly by the decay 
$(K_L \rightarrow \mu^+ \mu^-)_{SD}$
and the experimental value of $\varepsilon^\prime/\varepsilon$. Other 
observables constraining this coupling are $K^+ \to \pi^+ \nu \bar{\nu}$ and 
$\varepsilon_K$ \cite{BurasSilvestrini}.

The decay $K_L \rightarrow \mu^+ \mu^-$ is $CP$ conserving and in
addition to its short distance part, given by $Z$ penguins and box
diagrams, receives important contributions from the two photon
intermediate state which are difficult to calculate reliably.
Unfortunately, the separation of the short-distance part (similar to
$K^+ \rightarrow \pi^+ \nu \bar{\nu}$, free of hadronic uncertainties)
from the long-distance piece in the measured rate is rather
difficult. Therefore, the full branching ratio is generally written as
a sum of a dispersive and absorptive contributions, of which the
latter can be calculated using the data for $K_L \rightarrow \gamma
\gamma$. The dispersive contribution can be decomposed as a long
distance and a short distance part. Following
\cite{BurasSilvestrini,PichDumm}, we can write down,
\begin{eqnarray}
&Br\left( K_{L}\rightarrow \mu \overline{\mu }\right) _{SD}=6.32\times
10^{-3}\ \left[ C_{U2Z}\ \mbox{Re}\left( U_{sd}\right) 
\right.\nonumber \\
&\left. +\ \bar{\Delta}_c +
Y_{0}\left( x_{t}\right)\ \mbox{Re}\left( \lambda _{t}^{sd}\right)
\right]^{2}\ \leq\ 2.8\times 10^{-9},
\label{kmumu1}
\end{eqnarray}
where $C_{U2Z}=-(\sqrt{2} G_{F}M_{W}^{2}/\pi^{2})^{-1}\simeq - 92.7$, 
$\lambda _{i}^{ab}=V_{ia}^{\ast }V_{ib}$, $\bar{\Delta}_c=- 6.54
\times 10^{-5}$ is the charm quark contribution and $Y_{0}$ is the 
Inami-Lim \cite{I-L} function as defined in \cite{BurasH}. Experimentally, 
we must require this branching ratio to be 
$BR(K_L \rightarrow \mu^+ \mu^-)_{SD} \leq 2.8 \times 10^{-9}$ \cite{PichDumm}.

A second observable constraining $U_{sd}$ is
$\varepsilon^\prime/\varepsilon$.  Within the SM there is a
cancellation between QCD and electroweak penguin contributions
(dominated by $Z$ penguins) which suppresses this ratio. When new
physics enters into the game and if, as it is expected on general
grounds, it does not affect considerably the QCD contributions but
does so with the $Z$ penguins, the abovementioned cancellation does
not take place and significant deviations from the SM results (or
strong bounds on the new parameters) can be expected.  We now
decompose $\varepsilon^\prime/\varepsilon$ as follows,
\begin{eqnarray}
&\Frac{\varepsilon^{\prime}}{\varepsilon}=\ \beta_{U}\ C_{U2Z}\ \mbox{Im}\left(
U_{sd}\right)\ +\ \beta_{t}\ \mbox{Im}\left(\lambda _{t}^{sd}\right)\nonumber
\\ 
&\beta_{U}\ =\ [1.2\ -\ R_{s}\ r_{Z}\ B_{8}^{(3/2)}] \nonumber \\ 
&\beta_{t}\ =\ \beta _{U}\cdot C_{0}\ -\ 2.3\ +\ R_{s}\ [1.1\ r_{Z}\
B_{6}^{(1/2)}\ +\ 
\nonumber \\
&(1.0\ +\ 0.12\ r_{Z})\ B_{8}^{(3/2)}].
\label{epsilonprima}
\end{eqnarray}
where the first term comes from the $Z$ piece and the other one
contains all the remaining ones \cite{BurasSilvestrini}. It is worth
noting that unlike previous observables, here the theoretical errors
overwhelm the experimental precision. The main sources of
uncertainties lie in the parameters, $B_6^{(1/2)}$ and
$B_8^{(3/2)}$. The importance of these uncertainties is somehow
increased because of the cancellation we have mentioned.

Here, we take $R_s = 1.5 \pm 0.5$, $r_Z= 7.5 \pm 1$ and $B_8^{(3/2)}=
0.8 \pm 0.2$ \cite{BurasSilvestrini}.  The value of $B_{6}^{(1/2)}$
has caused some controversy in the literature because of the different
values obtained in different schemes.  We take two different values in order 
to illustrate the situation where new physics in the $s$--$d$ sector is needed
or not needed to reproduce the experimental value of 
$\varepsilon^\prime/\varepsilon$.  For set I, we use
$B_{6}^{(1/2)}=1.0\pm 0.2$ with the other parameters as given above, as in 
Refs.~\cite{BurasSilvestrini,BurasMart,Buras2001}
which comes from large $N_c$ calculations \cite{paschos} and lattice 
analysis, and this tends to favor the
presence of new physics in $U_{ds}$. In set II, we use
$B_{6}^{(1/2)}=1.3\pm 0.5$ in order to incorporate the predictions of
Refs.~\cite{Bertolini,PichPallante}, where inclusion of the correction
from final-state interactions in a chiral perturbation theory analysis
tends to favor the SM range. Still, there are other schemes where different
values for $B_6^{(1/2)}$ and $B_8^{(3/2)}$ are obtained 
\cite{ximo,sumrules,donoghue}. In fact, we have explicitly checked that 
with the values in \cite{ximo}, $B_6^{(1/2)}=2.5 \pm 0.4$ and 
$B_8^{(3/2)}=1.35 \pm 0.20$, the constraints on the FCNC couplings 
$U_{\alpha \beta}$ are still consistent with the results presented 
below. 

Numerically, we have for the central values
\begin{eqnarray}
\label{epsprim}
&\Frac{\varepsilon^{\prime}}{\varepsilon}=\ -7.8 \ C_{U2Z}\ \mbox{Im}\left(
U_{sd}\right)\ +\ \left\{\begin{array}{l} 6.2\\ 9.9 \end{array} \right.
\ \mbox{Im}\left(\lambda _{t}^{sd}\right)
\end{eqnarray}
with the parameters in set I and set II respectively and we calculate 
its errors with a Gaussian method. This observable has to reproduce the
experimentally obtained value of $(\varepsilon^\prime/\varepsilon)^{exp}=
(2.12 \pm 0.46) \times 10^{-3}$.

A theoretically cleaner constraint \cite{buchalla}, although less restrictive 
than the previous two constraints is provided by BR($K^+ \rightarrow
\pi^+ \nu \bar{\nu}$),
\begin{eqnarray}
\label{kpinu}
  {\rm BR}(K^+ \rightarrow \pi^+ \nu \bar{\nu}) &=& 1.55 \cdot 10^{-4}
\Biggl[ \biggl( C_{U2Z} Im \{U_{sd}\} + X_0(x_t) \cdot
  Im \lambda_t^{sd}\biggr)^2  \\
  &&+ \biggl( C_{U2Z} Re \{U_{sd}\} -2.11\cdot 10^{-4}  + X_0(x_t) Re
  \lambda_t^{sd}\biggr)^2\Biggr]\,\nonumber ,
\end{eqnarray}   
with $X_0(x_t)= C_0(x_t) -4B_0(x_t)$ a gauge invariant loop function
combination of boxes and vertices \cite{I-L,BurasH}.  This decay has
already been measured in the experiment E787 at BNL \cite{brook},
however, so far, a single event has been found and this is not enough
to provide a definite value for the branching ratio. Hence we take
here only the upper limit at $95 \%$ C.L.  $\mbox{BR}(K^+ \rightarrow
\pi^+ \nu \bar{\nu}) \leq 8.3 \times 10^{-10}$
\footnote{In Ref.~\cite{BBV}, the old value of this BR at 1 $\sigma$ was used,
therefore both a lower and an upper limit were considered}. 

Finally, we include also $\varepsilon _{K}$, whose leading-order expression 
is \cite{BurasH,BB}:
\begin{eqnarray}
&\varepsilon_{K} =
\Frac{e^{i\pi /4}G_{F}B_{K}F_{K}^{2}m_{K}}{6\Delta m_{K}}\ \mbox{Im}\left\{
-\eta_{t t} \left( U_{sd}\right)^{2} + 
\Frac{\alpha }{4\pi \sin ^{2}\theta_{W}}\right. \nonumber \\ 
&\left. \left[ 8\sum\limits_{i=c}^{t}\ \eta_{t i} Y_{0}\left( x_{i}\right)\
\lambda _{i}^{sd}U_{sd}\  - 
\sum\limits_{i,j=c}^{t}\ \eta_{i j} S_{0}(x_{i},x_{j})\ \lambda
_{i}^{sd}\lambda _{j}^{sd}\right] \right\}
\label{epsilon}
\end{eqnarray}
where $S_{0}$ is another Inami-Lim function \cite{BurasH} and the QCD
correction factors (which take into account short distance QCD effects)
are given as follows,
\begin{equation}
\eta_{cc}= 1.38 \pm 0,20, \;\;\;\;\;\; 
\eta_{tt}= 0.57 \pm 0.01\;\;\;\;\;\; 
\eta_{tc}= 0.47 \pm 0.04
\end{equation}
Here, contrary to Ref.~\cite{BurasSilvestrini}, the coefficient
$Y_{0}\left( x\right) $ of the linear term in $U_{ds}$ is
characteristic of the present model, therefore, in principle, the
irrelevance of $\varepsilon _{K}$ to constraint $U_{ds}$ is not fully
guaranteed.

At this point, it is important to emphasize that Eqs. (\ref{kmumu1}),
(\ref{epsilonprima}), (\ref{kpinu}) and (\ref{epsilon}) are completely
valid in the general model with $n$ additional VLdQs.

\begin{figure}
\begin{center}
\epsfxsize = 0.7 \linewidth
\epsffile{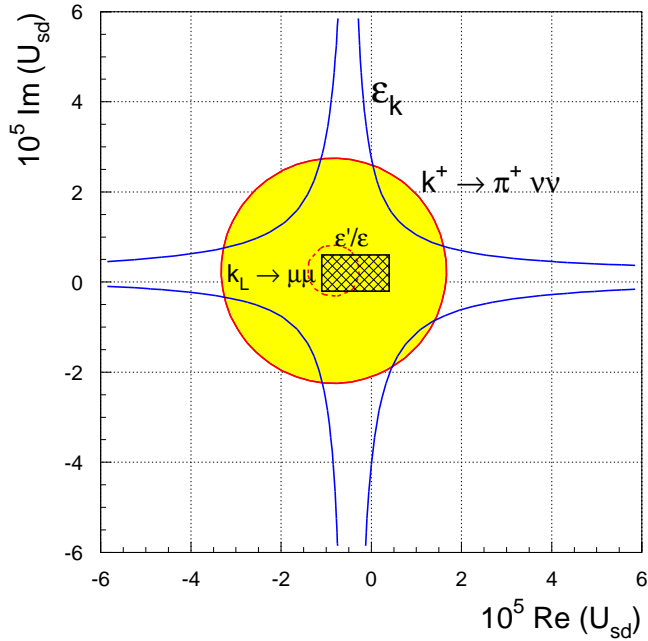}
\leavevmode
\end{center}
\caption{Effect of the constraints from $\varepsilon_K$, 
$K^+ \rightarrow \pi^+ \nu \bar{\nu}$, 
$K_L \rightarrow \mu^+ \mu^-$ and 
$\varepsilon^\prime/\varepsilon$ on the $U_{sd}$ FCNC coupling.}
\label{fig:constraints}
\end{figure}

In Fig.~\ref{fig:constraints} we present the effects of the different
constraints in the $U_{sd}$ coupling. As we can see in this figure,
the most efficient constraints are provided by $K_L \rightarrow \mu^+
\mu^-$, that constrains $\mbox{Re} (U_{sd})$, and
$\varepsilon^\prime/\varepsilon$, that bounds $\mbox{Im} (U_{sd})$ to
the rectangular box in the center. However, these constraints will be
difficult to improve due to the large hadronic
uncertainties. Similarly, the constraints from $\varepsilon_K$ are
very precise on the experimental side and the precision of this limit
on $U_{sd}$ is determined by the hadronic parameter $B_K = 0.85 \pm 0.15$
\cite{BurasH,Buras2001}.
Hence, it is not expected to improve largely in the near future. On
the other hand, the decay $K^+ \rightarrow \pi^+ \nu \bar{\nu}$ is
much cleaner from the theoretical point of view \cite{buchalla}. The
experimental value for the branching ratio, based in the single event
found so far is,
\begin{equation}
\mbox{BR}(K^+ \rightarrow \pi^+ \nu \bar{\nu})=
1.5^{+3.4}_{-1.2}\times 10^{-10},
\end{equation}
which gives an upper bound of,
\begin{equation}
BR(K^+ \rightarrow \pi^+ \nu \bar{\nu}) \leq 8.3 \times 10^{-10}\ \mbox{ at }\
95 \% \mbox{ CL}.
\end{equation}
 Unfortunately, it is clear from the errors that we cannot
obtain a lower bound on this BR at $80 \%$ CL, still we have included
in the figure a dashed circle showing the effect of a future lower
limit which would correspond to a hypothetical value of $8 \times
10^{-11}$ and would exclude the region to the left of the small
rectangle. Moreover, it is interesting to notice that improving the
upper bound to the level of $2 \times 10^{-10}$ would provide a bound
on the same level as the combined bounds from $K_L \rightarrow \mu^+
\mu^-$ and $\varepsilon^\prime/\varepsilon$. In fact, these values
could be reached after the analysis of the stored data from E-787 and
the sensitivity of the new experiment E-949 \cite{e949} (already
approved) will reach $10^{-11}$ in the next few years. Hence, this
decay will provide the most stringent and clean bounds on the $U_{sd}$
coupling in the near future, although at present the bounds are still
obtained from $K_L \rightarrow \mu^+ \mu^-$ and
$\varepsilon^\prime/\varepsilon$.
\begin{figure}
\begin{center}
\epsfxsize = 0.5 \linewidth
\epsffile{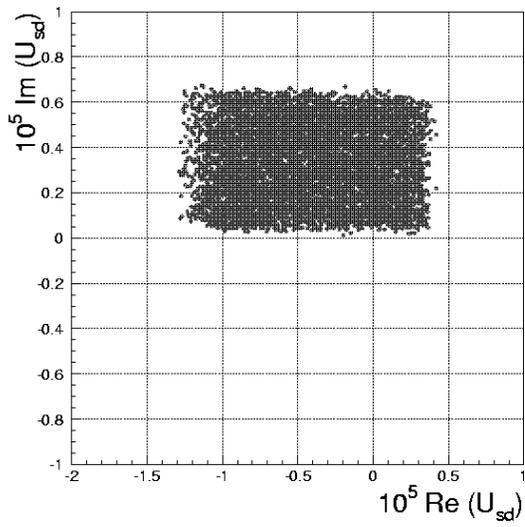}
\leavevmode
\end{center}
\caption{Phenomenological bounds on the FCNC coupling $U_{sd}$ with the
$\varepsilon^\prime/\varepsilon$ hadronic parameters in set I.}
\label{fig:lattice}
\end{figure} 
\begin{figure}
\begin{center}
\epsfxsize = 0.5 \linewidth
\epsffile{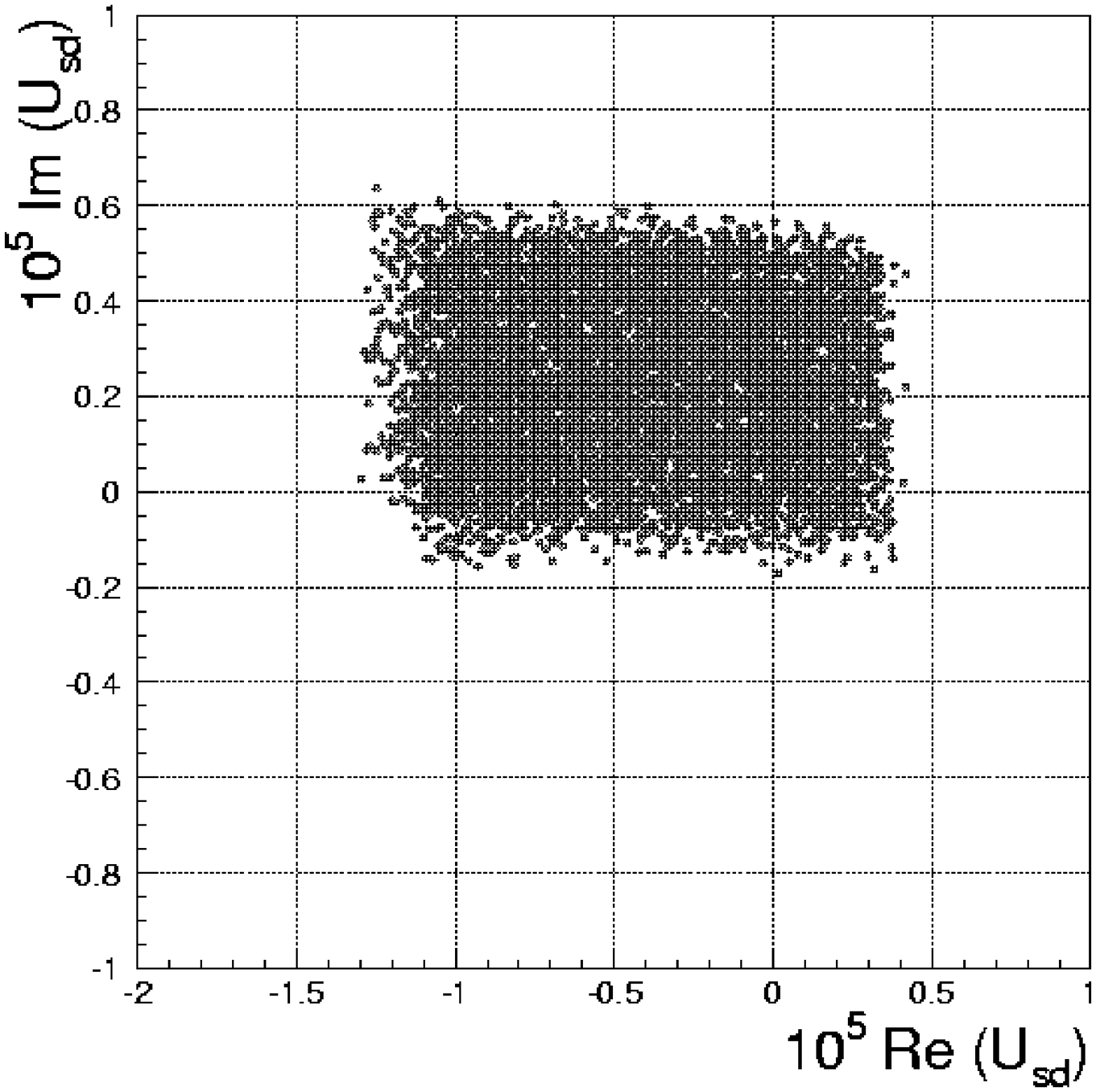}
\leavevmode
\end{center}
\caption{Phenomenological bounds on the FCNC coupling $U_{sd}$ with the
$\varepsilon^\prime/\varepsilon$ hadronic parameters from set II.}
\label{fig:chiral}
\end{figure} 

In Figs.~\ref{fig:lattice} and \ref{fig:chiral}, we present a scatter plot of
the allowed values of $U_{sd}$ in the general model with $n$ additional 
vector-like down quarks, although we must emphasize that
this allowed region does not change at all even if we go to the more 
restricted case of a single VLdQ. We impose all the constraints 
described above, with the $\varepsilon^\prime/\varepsilon$ parameters from 
set I and set II in Fig.~\ref{fig:lattice} and \ref{fig:chiral} respectively.  
In both approaches the bound for the real part is,
\begin{eqnarray}
-1.3 \times 10^{-5} < \mbox{Re}\{U_{sd}\}  < 4.0 \times 10^{-6}
\end{eqnarray}
This constraint is directly obtained from Eq.~(\ref{kmumu1}) with the
limit values of $\mbox{Re}(\lambda_t^{sd})$ which are
$\mbox{Re}(\lambda_t^{sd})_{max}= -4.9 \times 10^{-4}$ and
$\mbox{Re}(\lambda_t^{sd})_{min}= -2.1 \times 10^{-4}$ from
Eq.~(\ref{wolf4}) to ${\cal O}(\lambda^5)$ for $\phi=\pi,0$
respectively. Therefore, this implies that the additional correlations among 
the parameters in the 1--VLdQ model are irrelevant for the $U_{sd}$ bound.
 
The constraints on the imaginary part of this coupling depend slightly
on the adopted value for $B_{6}^{(1/2)}$. For set I,
with $B_{6}^{(1/2)}=1.0\pm 0.2$, we can see that $U_{sd}$ is
necessarily positive and does not reach the origin, indicating marginally the
need of new physics for $\varepsilon^{\prime}/\varepsilon$. The
allowed range is $ 1.0 \times 10^{-7} < \mbox{Im} \{U_{ds}\} < 6.5
\times 10^{-6}$. On the other hand, in set II, we get
an allowed area, $ - 1.5 \times 10^{-6} < \mbox{Im} \{U_{ds}\} < 6.0
\times 10^{-6}$, including the SM as one of the possible points in
agreement with the experimental results.  This bounds are directly
obtained from Eq.~(\ref{epsprim}) with
$\mbox{Im}(\lambda_t^{sd})_{max}= 1.4\times 10^{-4}$ and
$\mbox{Im}(\lambda_t^{sd})_{min}= 6 \times 10^{-5}$ assuming a $50 \%$
error, which corresponds to the Gaussian errors that we use in this
constraint. In view of these two
different options to calculate $\varepsilon^\prime/\varepsilon$, we
take a conservative bound on $ \mbox{Im}(U_{sd})$ including both
possibilities,
\begin{eqnarray}  
-1.5 \times 10^{-6} < \mbox{Im}\{U_{sd}\}  < 6.5 \times 10^{-6}
\end{eqnarray}
It is interesting to notice, that these bounds turn out to be much
more stringent than the direct bounds on FCNC couplings usually quoted
in the literature for VLdQ models \cite{nirrattazzi}.  This
improvement is mainly due to the inclusion of the SM contributions
that were completely neglected in the calculation of $\varepsilon_K$
in the bounds presented in \cite{nirrattazzi}, the improvement of the
experimental results in $(K_L \rightarrow \mu^+ \mu^-)_{SD}$, $K^+
\rightarrow \pi^+ \nu \bar{\nu}$, and to the inclusion of the
bound from $\varepsilon^\prime/\varepsilon$.  In this way, our bounds
basically agree with the general bounds in \cite{BurasSilvestrini}.
 
\subsection{B physics constraints}

In this section, we study the constraints on the $U_{bd}$ and $U_{bs}$ 
couplings from $B$ physics FCNC. In fact, the experimental information
in the $B$ system is improving rapidly with the new results
from $B$ factories and so, it is important to update the bounds on these 
FCNC couplings. In all the following processes, the choice of set I or set 
II to calculate $\varepsilon^\prime/\varepsilon$ has no 
relevant effects on the $U_{bd}$ and $U_{bs}$ couplings. Hence, we analyze 
the constraints making no distinction on the hadronic parameters 
used to calculate $\varepsilon^\prime/\varepsilon$.
In first place, we concentrate on the constraints from $CP$ 
conserving processes and then we add the information from the $B^0$ $CP$ 
asymmetries. 
\begin{figure}
\begin{center}
\epsfxsize =  \linewidth
\epsffile{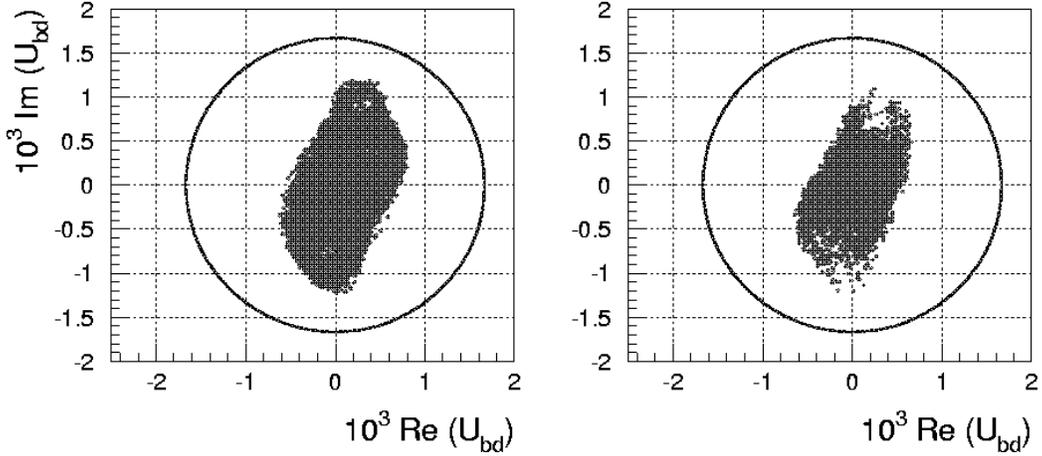}
\leavevmode
\end{center}
\caption{Constraints on the $U_{bd}$ coupling from $B_d$--$\bar{B}_d$
mixing in the $n$--VLdQ model (left) and in the 1--VLdQ model (right).
The circle shows the bound obtained from the $B \to X_d l^+ l^-$}
\label{fig:latticebd}
\end{figure} 
The main $CP$ conserving processes constraining $U_{bd}$ and $U_{bs}$ are 
$\mbox{BR}(B\rightarrow X_{d,s}l^+ l^-)$ and $\Delta M_{B_{d,s}}$.

From the upper bound on $\mbox{BR}(B\rightarrow X_{s}l^{+}l^{-}) \leq 4.2 
\times 10^{-5}$ \cite{cleo} and assuming $\mbox{BR}(B\rightarrow X_{d}l^{+}
l^{-}) \leq \mbox{BR}(B\rightarrow X_{s}l^{+}l^{-})$ we have 
\cite{BurasH,BHI},
\begin{eqnarray}
\Frac{\Gamma \left( B \to X_q l^+ l^-\right)}{\Gamma \left( B \to X_c e^+ 
\nu_e\right)}& =& \Frac{\alpha^2}{\pi^2 \sin^4 \theta_W}\ 
\Frac{\left| Y_{0}\left( x_{t}\right) \lambda _{t}^{bq}\ +\ 
C_{U2Z}\ U_{bq}\right|^2}{|V_{cb}|^2 f\left(m_c/m_b\right)}\times\nonumber \\
&&\left(2 \sin^4 \theta_W - \sin^2 \theta_W + \frac{1}{4}\right)
\leq 4 \times 10^{-4}
\end{eqnarray}
with $f(z)=( 1 - 8 z^2 + 8 z^6 -z^8 -24 z^4 \log z)$ a phase space
factor due to the mass of the charm quark. From here we get,
\begin{equation}
\left| Y_{0}\left( x_{t}\right) \lambda _{t}^{bq}\ +\ C_{U2Z}\
U_{bq}\right| <0.15.
\label{UBS}
\end{equation}
Replacing here $\lambda _{t}^{bd} \simeq A \lambda^3 (1 - \mu e^{-i
\phi}) + {\cal O}(\lambda^5)\simeq 8 \times 10^{-3}\ (1 - 0.4 e^{-i
\phi})$ with $Y_0 \simeq 1$, it is clear that the $\lambda _{t}^{bd}$
contribution can be safely neglected and we get a bound $|U_{bd}| \leq
1.7 \times 10^{-3}$. However, this is not true in the case of $\lambda
_{t}^{bs} \simeq - A \lambda^2 \simeq -0.040$ that shifts the
$|U_{bs}|$ constraint to a circle of radius $1.7 \times 10^{-3}$
centered in $(-0.040/92.7,0)$.
\begin{figure}
\begin{center}
\epsfxsize =  \linewidth
\epsffile{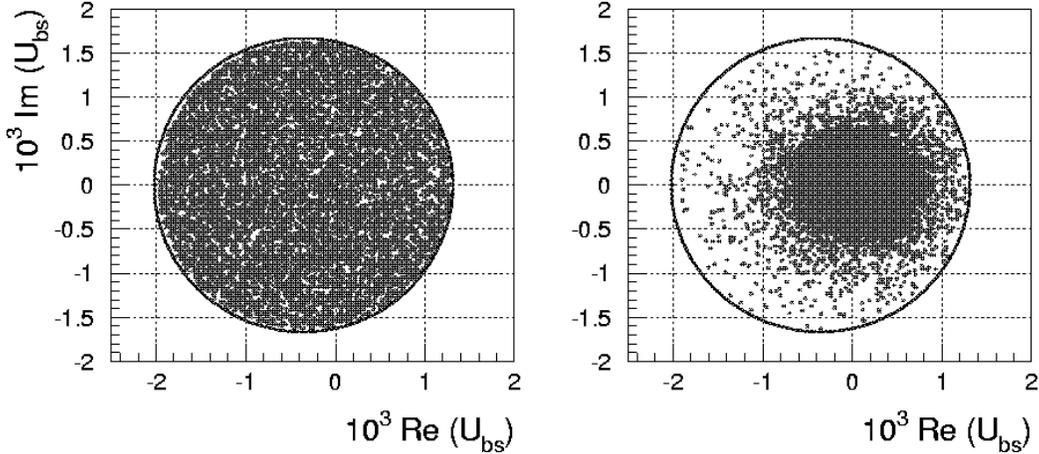}
\leavevmode
\end{center}\caption{Constraints on the $U_{bs}$ coupling from 
$B \to X_s l^+ l^-$ in the $n$--VLdQ model (left) and in
the 1--VLdQ model (right).}
\label{fig:latticebs}
\end{figure} 
 
Additionally, from $B$--$\bar{B}$ mixing  we have \cite{BBBV,BBV,BB} 
\begin{eqnarray}
&M^{B_q}_{12} = \Frac{G_{F}^{2}M_{W}^{2}
\eta_{B_{q}}B_{B_{q}}f_{B_{q}}^{2}m_{B_{q}}}{12\pi ^{2}} S_{0}\left(
x_{t}\right)\ {\lambda _{t}^{bq}}^{2} \Delta_{bq},\nonumber \\ 
&\Delta _{bq}\ =\ 1\ -\ 3.3\ \Frac{U_{bq}}{\lambda_{t}^{bq}}\ 
+\ 165\ \left( \Frac{U_{bq}} {\lambda_{t}^{bq}}\right)^{2},
\label{m12}
\end{eqnarray}
where the new parameters are defined in Ref.~\cite{BurasH}, and the mass 
difference $\Delta M_{B_{q}}= 2\  |M^{B_q}_{12}|$. 
The experimental values for these observables are 
$\Delta M_{B_{d}}=\left( 0.472\pm 0.017\right)
\times 10^{12}\ \mbox{s}^{-1}$ and $\Delta M_{B_{s}}>14.9\times
10^{12}\ \mbox{s}^{-1}$ at $95 \%$ C.L. \cite{stocchi,alilondon}. 

At this point, it is important to compare the bounds that can be
obtained from the mass difference and the semileptonic decay. From
$\mbox{BR}(B\rightarrow X_{d,s}l^+ l^-)$, we get a bound on $| \lambda
_{t}^{bq}Y_0 + C_{U2Z} U_{bq}|^2$ in Eq.~(\ref{UBS}), while from the
mass difference, neglecting the linear term, we obtain a constraint on
the combination $|(\lambda _{t}^{bq})^2 S_0 + 4 C_{U2Z}
U_{bq}^2|$. Hence, it is clear that the relative size of the tree
level FCNC with respect to the SM contribution is always bigger in the
semileptonic decay due to the presence of an additional factor
$C_{U2Z}\simeq -92.7$. Nevertheless, the experimental constraints from
$\mbox{BR}(B\rightarrow X_{d,s}l^+ l^-)$ are still much larger than
the typical SM prediction while the $\Delta M_{B_{d}}$ measurement is
already saturated by the SM contribution. This implies that the
$\Delta M_{B_{d}}$ constraint is more effective in the case of
$U_{bd}$ and both experiments give rise to constraints of the same order of magnitude, dominating at the end the $\Delta M_{B_{d}}$ bound. This
can be seen in Fig.~\ref{fig:latticebd}, where we show the allowed
region of the parameter space both in the general model with an
arbitrary number of VLdQs and in the minimal model with a single
VLdQ. In these figures, the constraint from $BR(B\rightarrow X_{d}l^+
l^-)$ is shown as a circle slightly shifted from the origin and a
radius of $1.7 \times 10^{-3}$. However, as we can see here, the
bounds on $U_{bd}$ are $|\mbox{Re} \{U_{bd} \}|\leq 0.7 \times
10^{-3}$ and $|\mbox{Im} \{U_{bd} \}|\leq 1.2 \times 10^{-3}$, which
are directly obtained from the $\Delta M_{B_{d}}$ constraint as the
envelope of the curves $|(\lambda _{t}^{bd})^2 S_0 + 4 C_{U2Z}
U_{bd}^2|$ with all the allowed values of $\lambda _{t}^{bd}$.  Notice
that already in this most constrained model, i.e. the model with a
single VLdQ, the $U_{bd}$ coupling is only bounded by these processes
and therefore we obtain the same bounds both in this case and in the
general case with $n$ VLdQ.
  
In the case of the $U_{bs}$ coupling the situation is completely
different.  In first place $\Delta M_{B_{s}}$ has not been measured
yet, and so only a lower bound on the mass difference is available
which is not useful to set a constraint on $U_{bs}$.  Moreover, it is
easy to see from Eq.~(\ref{m12}) that for similar values of $U_{bd}$
and $U_{bs}$, the FCNC effects on $\Delta M_{B_{s}}$ will be
suppressed by a factor $(\lambda_{t}^{bd}/\lambda_{t}^{bs})^2 \simeq
\lambda^2$ when compared with the effects on $\Delta M_{B_{d}}$. This
implies that in this model we cannot expect to observe the FCNC
effects in $B_s$--$\bar{B}_s$ mixing \footnote{This is not always true
in the presence of an up vector-like quark, as can be seen in
Ref.~\cite{BsBsbar}}.  Hence, this coupling is only constrained by the
upper bound on $\mbox{BR}(B\rightarrow X_{s}l^+ l^-)$.  In
Fig.~\ref{fig:latticebs}, we show the constraint from $BR(B\rightarrow
X_{s}l^+ l^-)$ as a circle slightly shifted from the origin, that
implies an upper bound of $|U_{bs}| \leq 2 \times 10^{-3}$
\cite{BHI,leptonic} both in the general model with $n$ VLdQs and in
the model with a single VLdQ. Similarly, the $b\to s \gamma$ branching
ratio provides constraints of the same order of magnitude
\cite{radiative}.

\begin{figure}
\begin{center}
\epsfxsize = \linewidth
\epsffile{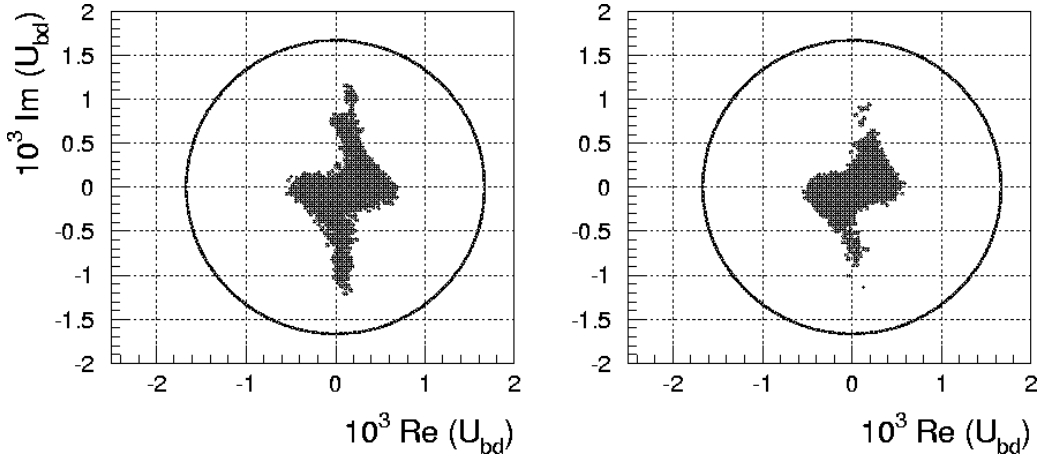}
\leavevmode
\end{center}\caption{Allowed region for the $U_{bd}$ coupling requiring a 
value of $a_{J/\psi}$ asymmetry to reproduce the world average at 1 $\sigma$
in the $n$--VLdQ model (left) and 1--VLdQ model (right).}
\label{fig:bdwav}
\end{figure}

Despite these strong constraints from $CP$ conserving processes, the $CP$ 
asymmetries in $B$ decays are still very effective to constrain the 
$U_{bd}$ coupling \cite{BEYALNir,EyalNir}. Recently, the arrival of the 
first measurements of the $B \to J/\psi K_S$ $CP$ asymmetry, $a_{J/\psi}$, 
from the $B$ factories has caused a great excitement in the high energy 
physics community.
\begin{equation}
\label{Bfactories}
\begin{array}{ll}
& \\
a_{J/\psi} & = \\ 
& \\
\end{array}
\left\{ 
\begin{array}{ll}
0.34\pm 0.20\pm 0.05 & (\mbox{BaBar \cite{babar}}) 
\\0.58 ^{+0.32+0.09}_{-0.34-0.10}
& \left( \mbox{Belle \cite{belle}}\right) \\ 
0.79^{+0.41}_{-0.44}
& \left( \mbox{CDF \cite{CDF}}\right)
\end{array}
\right.  \label{3ajk}
\end{equation} 
These values correspond to a world average of $a_{J/\psi}=0.51 \pm
0.18$, that can be compared with the SM expectations of $0.59 \leq
a_{J/\psi}^{SM}= \sin ( 2 \beta) \leq 0.82$ with $\beta=\mbox{arg} (-
V_{cd} V_{cb}^*/(V_{td} V_{tb}^*))$.  The errors are still too large
to draw any firm conclusion. In fact, if we take the world average at
$95 \%$ C.L. we do not get any improvement over the constraints from
$CP$ conserving processes. However, anticipating the improvement of
the experimental errors from $B$ factories, we take the world average
at 1 $\sigma$ level to see the effects on the $U_{bd}$ coupling.  From
Eq.~(\ref{m12}), it is straightforward to obtain $a_{J/\psi}$ as,
\begin{eqnarray}
&a_{J/\psi}=\sin \left(2 \beta -\mbox{arg}(\Delta_{bd}) \right).
\label{ajp}
\end{eqnarray}
Using the world average at 1 $\sigma$ as the experimentally allowed
range, we show in Fig.~\ref{fig:bdwav} the resulting region for the
$U_{bd}$ coupling.
As we can see here, the $U_{bd}$ allowed region is sizeably modified
from this constraint. The outer regions in the second and fourth
quadrants are reduced while the central region corresponding to the SM
remains filled; this situation represents an improvement over the
analysis presented in Ref.~\cite{EyalNir}. Note that the results are
essentially similar for the $n$--VLdQ and the 1--VLdQ cases. As we can
see here, with the world average for the $a_{J/\psi}$ $CP$ asymmetry
there is no need of new physics in the $U_{bd}$ sector.

\section{Discovery potential in $B$ factories}

As we have explained above, the present results in the $a_{J/\psi}$
asymmetry in Eq.~(\ref{3ajk}) are not precise enough to make a
definitive statement about the presence/absence of large new physics
contributions in $B$ decays. Still, these measurements, and especially
the BaBar value which is the most precise one, leave room for an
asymmetry considerably smaller than the standard model expectations.  
The SM range is certainly outside the $1\sigma $ BaBar
range but not outside the world average. This potential discrepancy is
at the origin of several papers \cite{smallajk,ajksusy} studying the
implications of a small $a_{J/\psi}$ in the search of new physics. The
papers in Ref.~\cite{smallajk} provide general parametrizations of new
physics contributions to $B$ mixing and decays. They essentially show
that this possible mismatch among the measured value and the SM
expectations may be due to the presence of new physics either directly
in $B$ physics or in $K$ physics modifying indirectly the unitarity
triangle fit. However, these papers do not provide a definite new
physics example fulfilling this task. On the other hand, the two
papers in Ref.~\cite{ajksusy} analyze SUSY models, where no sizeable
effects in $B$ mixing and decays are generally expected and the
unitarity triangle fit can be modified only through new SUSY
contributions to $K$ physics. In the following, we show that a model
with tree level FCNC from the mixing with vector-like quarks would be
a natural candidate for a model with clear deviations from the SM
expectations in $CP$ asymmetries, satisfying simultaneously all other
experimental constraints.

\begin{figure}
\begin{center}
\epsfxsize = \linewidth
\epsffile{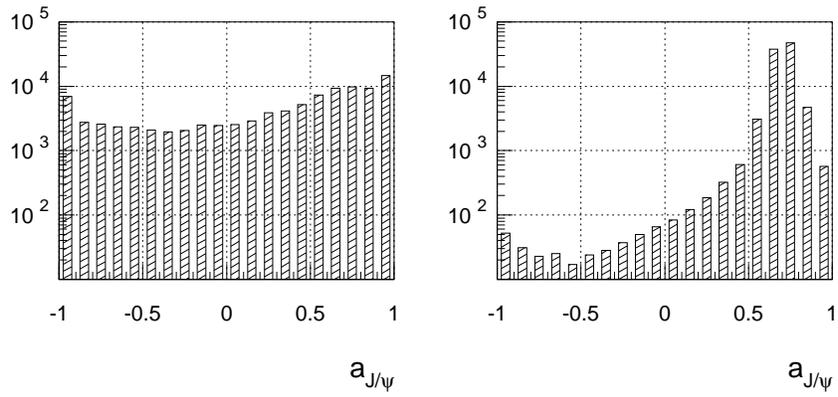}
\leavevmode
\end{center}\caption{Distribution of the values of $a_{J/\psi}$ for 95000 
events in the $n$--VLdQ (left) and in the 1--VLdQ (right).}
\label{fig:barrasym}
\end{figure} 
From the expression of $a_{J/\psi}$ in Eqs.(\ref{m12}) and
(\ref{ajp}), with $|\lambda_{t}^{bd}| \in [5 \times 10^{-3}, 1.2
\times 10^{-2}]$ and $|U_{bd}| \lsim 10^{-3}$, it is clear that
$\mbox{arg}(\Delta_{bd})$ can be very large.  Indeed, in the general
model with an arbitrary number of VLdQ, there is no further influence
on $U_{bd}$ from constraints on other elements of the mixing
matrix. In fact, only the inequalities in Eq.~(\ref{ineq}) can
restrict this coupling, however, given the bounds on $D^2_i$, they
are almost inoperative in this case. Hence, all the points inside the
allowed contour in Fig.\ref{fig:latticebd} are equally probable and
any value of the asymmetry is possible with a comparable
probability.
Still, the especial geometry of this allowed region and
the preferred orientation of $\lambda^t_{bd}$ slightly favors the SM
range over other values. Furthermore, even in a more constrained
model, as in the model with a single VLdQ large departures from the SM
range are possible. The expected values for the asymmetry both in the
$n$ VLdQs and in the 1 VLdQ model are shown in Fig.~\ref{fig:barrasym}
as an histogram of the distribution of 95000 events satisfying all the
constraints in the previous sections. As expected from the above
discussion, in the general case all possible values of the asymmetry
have similar probability. On the other hand, in the model with a
single VLdQ, this distribution is clearly peaked on the SM
range. Moderate deviations are rather frequent and larger departures
are more rare but still possible for any value of the asymmetry.
\begin{figure}
\begin{center}
\epsfxsize = 0.5 \linewidth
\epsffile{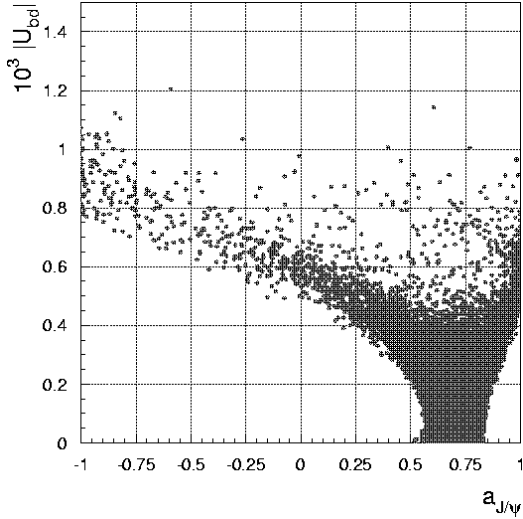}
\leavevmode
\end{center}\caption{Correlation between the values of $a_{J/\psi}$ and 
$|U_{bd}|$ in the model with a single VLdQ.}
\label{fig:asymzbd}
\end{figure}
\begin{figure}
\begin{center}
\epsfxsize = 0.5 \linewidth
\epsffile{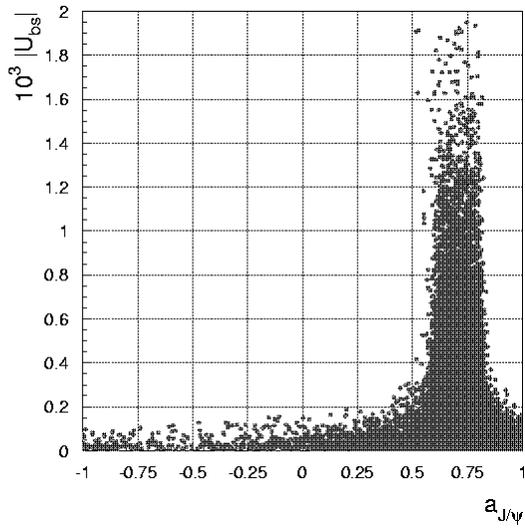}
\leavevmode
\end{center}\caption{Correlation between the values of $a_{J/\psi}$ and 
$|U_{bs}|$ in the model with a single VLdQ.}
\label{fig:asymzbs}
\end{figure}

\begin{figure}
\begin{center}
\epsfxsize = \linewidth
\epsffile{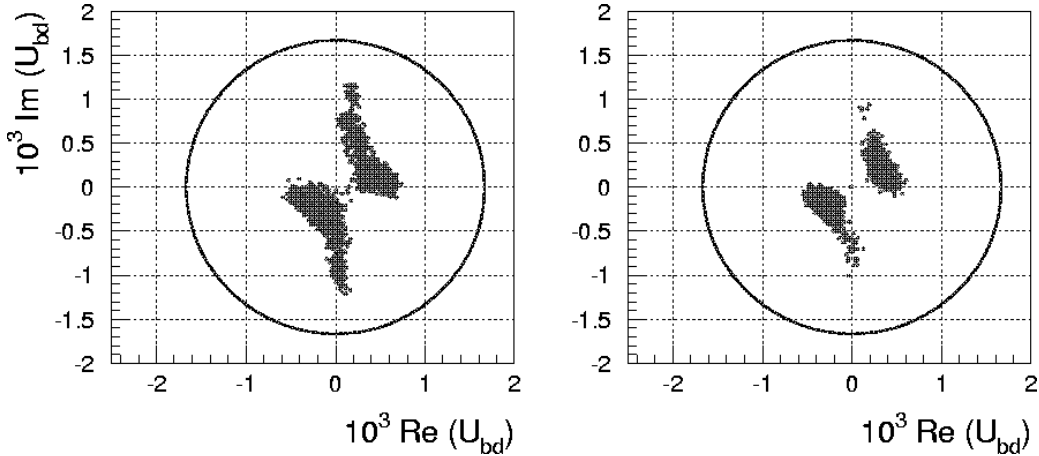}
\leavevmode
\end{center}\caption{Allowed region for the $U_{bd}$ coupling consistent 
with the BaBar result for the $a_{J/\psi}$ asymmetry at 1 $\sigma$
in the $n$--VLdQ (left) and in the 1--VLdQ (right).}
\label{fig:bdbb}
\end{figure}

\begin{figure}
\begin{center}
\epsfxsize = \linewidth
\epsffile{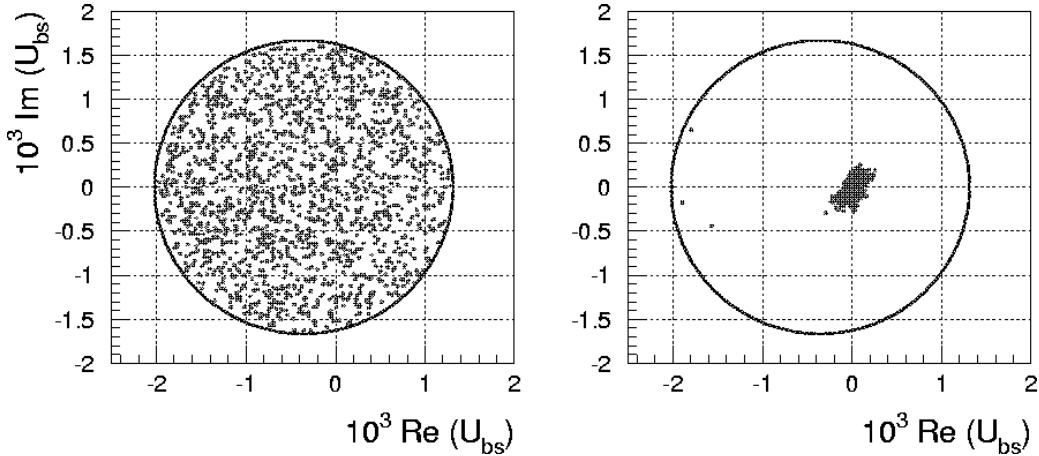}
\leavevmode
\end{center}\caption{Allowed region for the $U_{bs}$ coupling consistent 
with the BaBar result for the $a_{J/\psi}$ asymmetry at 1 $\sigma$
in the $n$--VLdQ (left) and in the 1--VLdQ (right).}
\label{fig:bsbb}
\end{figure}

In Figs.~\ref{fig:asymzbd} and \ref{fig:asymzbs}, we show, in the 1
VLdQ model the correlation of the possible values of the asymmetry
with $|U_{bd}|$ and $|U_{bs}|$ respectively. Here, we see that the
range $0.55 \lsim a_{J/\psi} \lsim 0.85$ corresponding to the SM
expected range concentrates most of the events and can be reproduced
with $|U_{bd}|=0$ and any allowed value of $|U_{bs}|$.  However, for
different values of the asymmetry there is a clear correlation between
$a_{J/\psi}$ and the minimum value of $|U_{bd}|$ required to obtain
this asymmetry. For instance to obtain an asymmetry below $0.5$, a
$|U_{bd}|\geq 2 \times 10^{-4}$ is needed. This required minimum is
also true in the general model with an arbitrary number of
VLdQs. Similarly, we see in Fig.~\ref{fig:asymzbs} that these large
values of $U_{bd}$ correspond to low values of $U_{bs}$ \cite{BBV},
however it is important to emphasize that this correlation is only
true in the minimal model, but not in a model with an arbitrary number
of VLdQs. To see this we show in Figs.~\ref{fig:bdbb} and
\ref{fig:bsbb} both in the $n$--VLdQ and in the 1--VLdQ models
an scatter plot  of $\mbox{Re}
(U_{bd})$ versus $\mbox{Im} (U_{bd})$ and $\mbox{Re} (U_{bs})$ versus
$\mbox{Im} (U_{bs})$ for an asymmetry corresponding to the BaBar
result at $1 \sigma$, $0.14 \leq a_{J/\psi} \leq0.54$ \cite{babar}.
In this Fig.~\ref{fig:bdbb} we see that the great majority of the
allowed points are in the range $1 \times 10^{-4}\ (2 \times
10^{-4})\lsim \left| U_{bd}\right| \lsim 1.2\times 10^{-3}$, in the
$n$-VLdQ (1--VLdQ) model, i.e. a large, non-vanishing $U_{bd}$
coupling is required to reproduce the BaBar asymmetry. In particular,
this means that, within this model, a low CP asymmetry implies the
presence of new physics in $b$--$d$ transitions. This conclusion would be
unchanged in the general model. On the other hand, we see that, for
these points, in the 1-VLdQ model, the coupling $U_{bs}$ is always
restricted to the range $|U_{bs}|\lsim 3 \times 10^{-4}$; hence all
the allowed points have simultaneously high $\left|U_{bd}\right|$ and
low $\left| U_{bs}\right|$.  Indeed, it is easy to obtain in the
1--VLdQ model, from Eq.~(\ref{min4}), the relation $U_{bd}U_{bs}^{*}=-
U_{sd} D_b^{2}$. The region in the $U_{ds}$ plane does not change with
the inclusion of the $a_{J/\psi}$ constraint, and then we still have,
$|U_{sd}|\lsim 6 \times 10^{-6}$ and $D_b^2 \lsim 0.009$. Taking into
account that a low $a_{J/\psi}$ requires $|U_{bd}| \geq 2 \times
10^{-4}$, this clearly implies an absolute upper bound, $|U_{bs}|\lsim
3 \times 10^{-4}$.  Therefore, for this set of points, we can not
expect a new-physics contribution in the $b \to s$ transition.
However, this relation is not valid in the $n$--VLdQ model and
therefore this correlation is lost as can be seen in
Fig.\ref{fig:bsbb}.
 
Also in Fig.~\ref{fig:bdbb}, we find a few points ($\simeq 0.1 \%$
of the points) which have simultaneously $|U_{bs}|\gsim 1 \times
10^{-3}$ and $|U_{bd}|\lsim 3 \times 10^{-5}$.  This second class
of points is only possible in the vicinity of the SM and they
disappear if the value of the asymmetry is reduced to
$a_{J/\psi}\lsim 0.52$  \footnote{Still, it is important to emphasize that
these points also require the presence of new physics in $B$
decays.}. 

Therefore, we can conclude that in a model with a single VLdQ a low $CP$ 
asymmetry $a_{J/\psi} \leq 0.52$ implies $|U_{bd}| \geq 2 \times 10^{-4}$
and simultaneously $|U_{bs}| \leq 3 \times 10^{-4}$. On the contrary, in
a general model with an arbitrary number of VLdQs, we still have
$|U_{bd}| \geq 2 \times 10^{-4}$, but this has no influence on the
$U_{bs}$ coupling, and it is only constrained by the $B \to X_s l \bar{l}$
branching ratio.

At this point, it is very interesting to examine the predicted
branching ratios of the decays $B\to X_{d,s} l \bar{l}$ for this set
of points.  From Fig.~\ref{fig:bdbb}, where we have included the
circle corresponding to the experimental bounds in these decays, it is
clear that we can also expect a very large contribution to $B\to X_{d}
l \bar{l}$. In this case, the branching ratios for the $X_d$ decays
are strongly enhanced from the SM prediction, reaching values of $1.0
\times 10^{-6} \leq BR\left(B\rightarrow X_{d}l^{+}l^{-}\right) \leq
1.8 \times 10^{-5}$ and $6.0 \times 10^{-5} \leq BR\left(B\rightarrow
X_{d}\nu \bar{\nu}\right) \leq 1.0 \times 10^{-4}$. While, on the
other hand, the low values of $U_{bs}$ imply that the $X_s$ decays
remain roughly at the SM value. Conversely, in the points with small
$U_{bd}$ and large $U_{bs}$, there is no sizeable departure from the
SM expectations in $B\to X_{d} l \bar{l}$ and the $X_s$ decays are now
close to the experimental upper range. Namely, we obtain, for the
points to the right of Fig.~\ref{fig:bsbb}, with
$Re\left(U_{bs}\right) \simeq 1 \times 10^{-4}$, $BR\left(B\rightarrow
X_{s}l^{+}l^{-}\right) \simeq 2.7 \times 10^{-5} $ to be compared with
the experimental upper bounds of $BR\left(B\rightarrow
X_{s}l^{+}l^{-}\right)\leq 4.2 \times 10^{-5}$.  However, this
possibility is only marginal in the $1 \sigma$ BaBar range for the
minimal model with a single VLdQ. In any case in the general model,
the possibility of large $U_{bs}$ couplings is again open. For
analysis of the phenomenological effects of this coupling in radiative
$b\to s \gamma$ decays, $B_s$--$\bar{B}_s$ mixing and rare decays in a
slightly different context, see
Refs.~\cite{BsBsbar,leptonic,radiative}.

\section{Conclusions}

In this work, we have updated the constraints on tree level FCNC
couplings in the framework of a theory with $n$ isosinglet vector-like
down quarks. The inclusion of the constraints from $(K_L\to \mu^+
\mu^-)_{SD}$, $\varepsilon^\prime/\varepsilon$, $\varepsilon_K$ and $K
\to \pi \nu \bar{\nu}$ has allowed to improve sizeably the bound on
the $U_{sd}$ coupling. The precise range of allowed values for this
coupling depends strongly on the hadronic input in the calculation of
$\varepsilon^\prime/\varepsilon$. Our summary in
Table~\ref{tab:constr} includes the main theoretical approaches
\cite{BurasSilvestrini,BurasMart,Buras2001,paschos,Bertolini,PichPallante,
ximo,sumrules,donoghue}. In the near future, $K^+ \to \pi^+ \nu
\bar{\nu}$ will be quite useful to further constrain $U_{sd}$ without
large theoretical uncertainties.

We have calculated the constraints on $U_{bd}$ and $U_{bs}$ from
$\Delta M_{B_d}$, $B \to X_{s,d} l \bar{l}$ and the $B \to J/\psi K_S$
$CP$ asymmetry. In Table~\ref{tab:constr}, we summarize our results on
these bounds without the additional constraint from the $B \to J/\psi
K_S$ $CP$. Note that all these bounds on $U_{bq}$ are independent of the 
hadronic inputs in $\varepsilon^\prime/\varepsilon$.
 \begin{table}
\begin{center} 
\begin{tabular}{||c|c|c||} \hline \hline Re $\{ U_{sd}
 \}$ & $\geq - 1.3 \times 10^{-5}$ & $\leq 4.0 \times 10^{-6}$\\
 \hline Im $\{ U_{sd} \}$ &$\geq - 1.5 \times 10^{-6}$ & $\leq 6.5
 \times 10^{-6}$\\ \hline $ |\mbox{Re} \{ U_{bd} \}|$
 &\multicolumn{2}{c||}{$\leq 7 \times 10^{-4}$}\\ \hline $ |\mbox{Im}
 \{ U_{bd} \}|$ &\multicolumn{2}{c||}{$\leq 1.2 \times 10^{-3}$}\\
 \hline $|U_{bs}|$ &\multicolumn{2}{c||}{$\leq 2 \times 10^{-3}$}\\
 \hline \hline 
\end{tabular} 
\caption[]{Constraints on the tree level
 FCNC couplings from rare processes in the $K$ and $B$ systems.}
\label{tab:constr} 
\end{center} 
\end{table} 
In addition, we have shown that the $B \to J/\psi K_S$ $CP$ asymmetry
 is especially sensitive to the presence of a $U_{bd}$ FCNC
 coupling. In this regard, we have shown that the 1 $\sigma$ value of
 the world average is already able to exclude half of the allowed
 region for the $U_{bd}$ coupling.  However, due to the still large
 experimental errors this constraint has a small effect at $95 \%$
 C.L.  Assuming a low value of the $a_{J/\psi}$ asymmetry below $\sim
 0.52$ (following BaBar range at 1 $\sigma$) we have shown that, in
 models with $n$ VLdQs, tree-level FCNC in the $b$--$d$ sector are
 mandatory. The FCNC coupling $U_{bd}$ is required to be greater than
 $1 \times 10^{-4}$, rising to $2 \times 10^{-4}$ in the 1--VLdQ
 model.
The $U_{bs}$ coupling in the general model is bounded by $2 \times 10^{-3}$,
although for the 1--VLdQ model the bound goes down to $3 \times 10^{-4}$.

Therefore, a clear favorable scenario for these models would be to
find simultaneously a low $a_{J/\psi}$ (below $\sim 0.5$) and a large
$\mbox{BR}( B_d \to X_d l \bar{l})$, at least one order of magnitude
bigger than the SM expectations. Values of $\mbox{BR}( B_d \to X_s l
\bar{l})$ close to the present experimental bounds are still possible
in the $n$--VLdQ model. Nevertheless, in the 1--VLdQ model these BR
are expected to be similar to the SM values.  In summary, this
$a_{J/\psi}$ $CP$ asymmetry, together with the rare decays $B \to X_{s,d}
l \bar{l}$ and $K \to \pi \nu \bar{\nu}$ are the best options to
further constraint the FCNC tree level couplings or to discover the
presence of vector-like quarks in the low energy spectrum as suggested
by GUT theories or models of large extra dimensions at the TeV scale.

\section*{Acknowledgements} 
The work of F.B. and O.V. was partially supported by the Spanish CICYT
AEN-99/0692. O.V.  acknowledges financial support from a Marie Curie
EC grant (HPMF-CT-2000-00457).

\end{document}